\documentclass{article}

\usepackage{PRIMEarxiv}

\usepackage[utf8]{inputenc} 
\usepackage[T1]{fontenc}    
\usepackage{hyperref}       
\usepackage{url}            
\usepackage{booktabs,caption}       
\usepackage{amsfonts}       
\usepackage{nicefrac}       
\usepackage{microtype}      
\usepackage{lipsum}
\usepackage{fancyhdr}       
\usepackage{graphicx,xcolor}       
\usepackage{hyperref, comment}
\graphicspath{{media/}}     

\title{The Effectiveness of a Virtual Reality-Based Training Program for Improving Body Awareness in Children with Attention Deficit and Hyperactivity Disorder
\thanks{This study has been approved by the IRB Approval Committee at VRapeutic Inc. and the Research Ethics Approval Committee of the Electronics and Communications Engineering Department at the Arab Academy for Science and Technology (VRAP.MO/06).} 
}

\author{
  Aya Abdelnaem El-Basha \\
  Faculty of Education for Early Childhood \\
  Damanhour University \\
  \texttt{ayaabdelnaiem@gmail.com} \\
  \And
  Ebtsam ELSayed Mahmoud ELSayes \\
  Faculty of Education for Early Childhood \\
  Damanhour University \\
  \texttt{ebtsamelsayes10@kgr.dmu.edu.eg} \\
  \And
  Ahmad Al-Kabbany \\
  VRapeutic Inc. \\
  Multimedia Interaction and Communication Lab \\
  Wearables, Biosensing, and Biosignal Processing Research Lab \\
  Arab Academy for Science and Technology \\
  \texttt{alkabbany@ieee.org, alkabbany@aast.edu} \\
}

\begin{document}
\maketitle

\begin{abstract}
This study investigates the effectiveness of a Virtual Reality (VR)-based training program in improving body awareness among children with Attention Deficit Hyperactivity Disorder (ADHD). Utilizing a quasi-experimental design, the research sample consisted of 10 children aged 4 to 7 years, with IQ scores ranging from 90 to 110. Participants were divided into an experimental group and a control group, with the experimental group receiving a structured VR intervention over three months, totaling 36 sessions. Assessment tools included the Stanford-Binet Intelligence Scale (5th Edition), the Conners Test for ADHD, and a researcher-prepared Body Awareness Scale.  

The results indicated statistically significant differences between pre-test and post-test scores for the experimental group, demonstrating the program's efficacy in enhancing spatial awareness, body part identification, and motor expressions. Furthermore, follow-up assessments conducted one month after the intervention revealed no significant differences from the post-test results, confirming the sustainability and continuity of the program’s effects over time. The findings suggest that immersive VR environments provide a safe, engaging, and effective therapeutic medium for addressing psychomotor deficits in early childhood ADHD.
\end{abstract}


\section{Introduction}
\label{sec:Intro}
Attention Deficit Hyperactivity Disorder (ADHD) is recognized as one of the most prevalent neurodevelopmental disorders among children. Global estimates from the Diagnostic and Statistical Manual of Mental Disorders (DSM-5) suggest a prevalence of approximately 5\%, though recent reports indicate figures as high as 11.4\% in certain populations \cite{danielson2018prevalence}. In early childhood settings, such as kindergartens, the incidence can reach 12\%, serving as a significant predictor of long-term developmental challenges.

Beyond the core symptoms of inattention and impulsivity, children with ADHD frequently exhibit a significant impairment in body awareness \cite{yu2024effect}. This psychomotor component involves the child’s ability to perceive their body’s position in space, identify body parts, and understand their relationship to the surrounding environment. Such deficits manifest as clumsy movements, difficulty with balance, and poor spatial organization, which negatively impact fine motor skills, visual-motor coordination, and overall academic and social integration.

Traditional therapeutic approaches often struggle to maintain the engagement of children with ADHD, who are prone to distraction by minor external stimuli. Furthermore, prolonged exposure to traditional, passive screens has been observed to weaken learning motivation and limit the opportunities for developing spatial and body awareness. There is a critical need to transform these technological interactions from passive consumption into active, therapeutic engagement.

Virtual Reality (VR) has emerged as a promising technological medium for intervention, offering immersive 3D environments that simulate physical presence. VR allows for the creation of safe, interactive, and controlled settings where children can practice motor tasks that would be difficult or risky in the real world \cite{konacc2024enhancing, corrigan2023immersive}. By providing immediate sensory feedback, VR can enhance attention and promote self-regulation through simulated real-life scenarios \cite{youssef2024telehealth,gaafer2024immersive}.

Despite the potential of VR, there is a notable scarcity of research focusing on body awareness for children with ADHD using this technology. Most existing studies focus on academic performance or general attention, leaving the foundational psychomotor skill of body awareness underserved. This study addresses the observed inability of children with ADHD to plan movements, maintain balance, and perceive their physical boundaries.

The primary objective of this research is to evaluate the effectiveness of a specialized VR-based training program in improving body awareness among children aged 4 to 7 years. The study also examines the sustainability of these improvements through longitudinal follow-up. By proposing this framework, we contribute an evidence-based technological solution to the field of early childhood special education, aiming to improve the diagnostic and rehabilitative outcomes for this population. The primary contributions of this study are as follows:

\begin{enumerate}
  \item Novel Therapeutic Framework: We present an evidence-based Virtual Reality (VR) training framework specifically tailored for children with ADHD in the early childhood stage (ages 4–7) to address psychomotor and body awareness deficits.

  \item Psychometric Validation: The study provides a researcher-developed and statistically validated "Body Awareness Scale" designed to measure three critical dimensions: body part identification, spatial positioning, and motor expressions.

  \item Empirical Evidence of VR Efficacy: We provide quantitative proof that immersive 3D environments significantly outperform traditional methods in enhancing spatial perception and motor planning for neurodivergent children.

  \item Sustainability Insights: This research demonstrates the longitudinal effectiveness of VR interventions, proving that psychomotor gains are maintained post-treatment, suggesting lasting neuroplastic adaptations.

  \item Localization of Digital Therapeutics: This work contributes to the sparse body of research in Arabic-speaking countries regarding the integration of "Embodied Technology" in special education and rehabilitation.
\end{enumerate}

The rest of this article is summarized as follows: Section 2 examines the theoretical foundations of body awareness in ADHD and surveys the current state of Virtual Reality in pediatric rehabilitation. Section 3 details the quasi-experimental design, participant demographics, the structure of the 36-session VR program, and the psychometric instruments used for data collection. Section 4 presents the statistical analysis of the intervention's impact and provides an in-depth discussion of the findings, including the study's limitations. Section 5 summarizes the research highlights, discusses practical implications for educators and therapists, and outlines avenues for future technological development.

\section{Related Work}
\label{sec:LitRev}
Body awareness is a foundational psychomotor skill encompassing the ability to identify body parts, understand their functions, and perceive their position and movement in space. According to the literature, this awareness is not innate but develops through distinct stages \cite{meachon2025motor}. The present study targets the "Perceptual Differentiation" stage (ages 3 to 7), a critical period where children begin to consciously distinguish their body boundaries from the environment. Successful development during this stage results in a "Body Schema"—a mental map that serves as the primary pillar for motor organization and environmental interaction \cite{lelong2021effective}.

Children diagnosed with ADHD frequently exhibit deficits that extend beyond cognitive inattention into the realm of motor planning. Research indicates that these children often display "spatial neglect," characterized by a disproportionate decline in spatial attention and alertness. This neglect manifests as a failure to recognize the left side of the visual field or a lack of coordination in bilateral body movements. Such psychomotor impairments are strongly correlated with delays in academic skills, as the child lacks the visual-motor integration necessary for tasks like writing, drawing, and maintaining social physical boundaries \cite{diaz2025sex, kaplan2025effect}.

Virtual Reality (VR) represents a paradigm shift in digital therapeutics by providing immersive, three-dimensional environments that simulate physical presence. Unlike traditional 2D screens, which the researcher notes can decrease learning motivation in ADHD populations, VR provides a "quasi-realistic" interactive experience \cite{vinci2026augmented}. By engaging the child in a multisensory feedback loop, VR encourages "Embodied Learning," where the child must physically navigate the virtual space to achieve goals. This interactive nature has been shown to improve both attention and self-regulation by reducing external distractors and providing a safe, controlled environment for individualized progress \cite{single2025effectiveness, maya2025immersive, mohamed2026toshfa}.

Despite the documented benefits of VR in global clinical settings, there is a significant lack of research addressing body awareness within the Arabic educational framework. Most interventions for ADHD focus exclusively on behavioral or academic outcomes, often ignoring the underlying psychomotor deficiencies that hinder a child’s foundational development. This study seeks to bridge this gap by implementing a structured VR program specifically aimed at improving body awareness through simulated spatial challenges \cite{wang2025impact}.

\section{Methodology}
\subsection{Research Design and Participants}
\begin{figure}
  \centering
  \includegraphics[width=\textwidth]{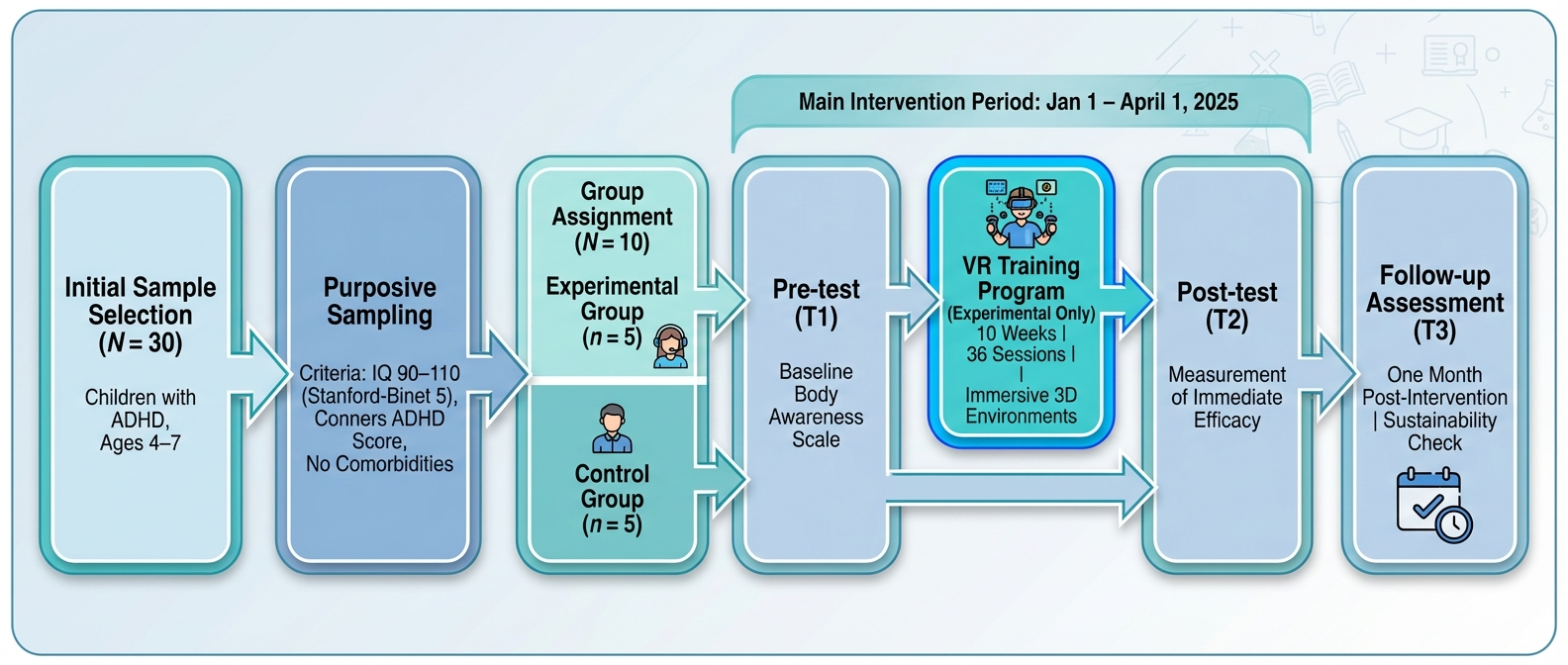}
  \caption{The experimental timeline and workflow of the proposed research}
  \label{fig:exp_timeline}
\end{figure}

This study utilized a quasi-experimental approach, specifically a two-group design (experimental and control) featuring pre-test, post-test, and follow-up assessments. The research was conducted at an early intervention center in Alexandria, Egypt, under the supervision of the Ministry of Social Solidarity. From an initial screening of 30 children, a final sample of 10 children aged 4 to 7 years was purposively selected. To ensure a high degree of internal validity, the researcher applied rigorous inclusion criteria: participants were required to have a confirmed diagnosis of ADHD via the Conners Rating Scale and an IQ score between 90 and 110 on the Stanford-Binet Scale, 5th Edition. Children with comorbid intellectual, sensory, or physical disabilities were excluded from the study to isolate the impact of the VR intervention on psychomotor body awareness. The study's timeline and workflow is dpeicted in Fig.~\ref{fig:exp_timeline}.

\subsection{Group Equivalence and Control}
The final cohort was divided into an experimental group ($n=5$) and a control group ($n=5$). Before the intervention, the researcher verified the homogeneity of the groups across several variables to prevent confounding results. Statistical analysis using the Mann-Whitney U test confirmed that there were no significant differences ($p > 0.05$) between the two groups in terms of chronological age, socioeconomic and cultural status, IQ scores, or baseline body awareness performance. This equivalence established a stable baseline for measuring the effectiveness of the independent variable—the virtual reality program—against the dependent variable of body awareness.

\subsection{Psychometric and Diagnostic Instruments}
\begin{figure}
  \centering
  \includegraphics[width=\textwidth]{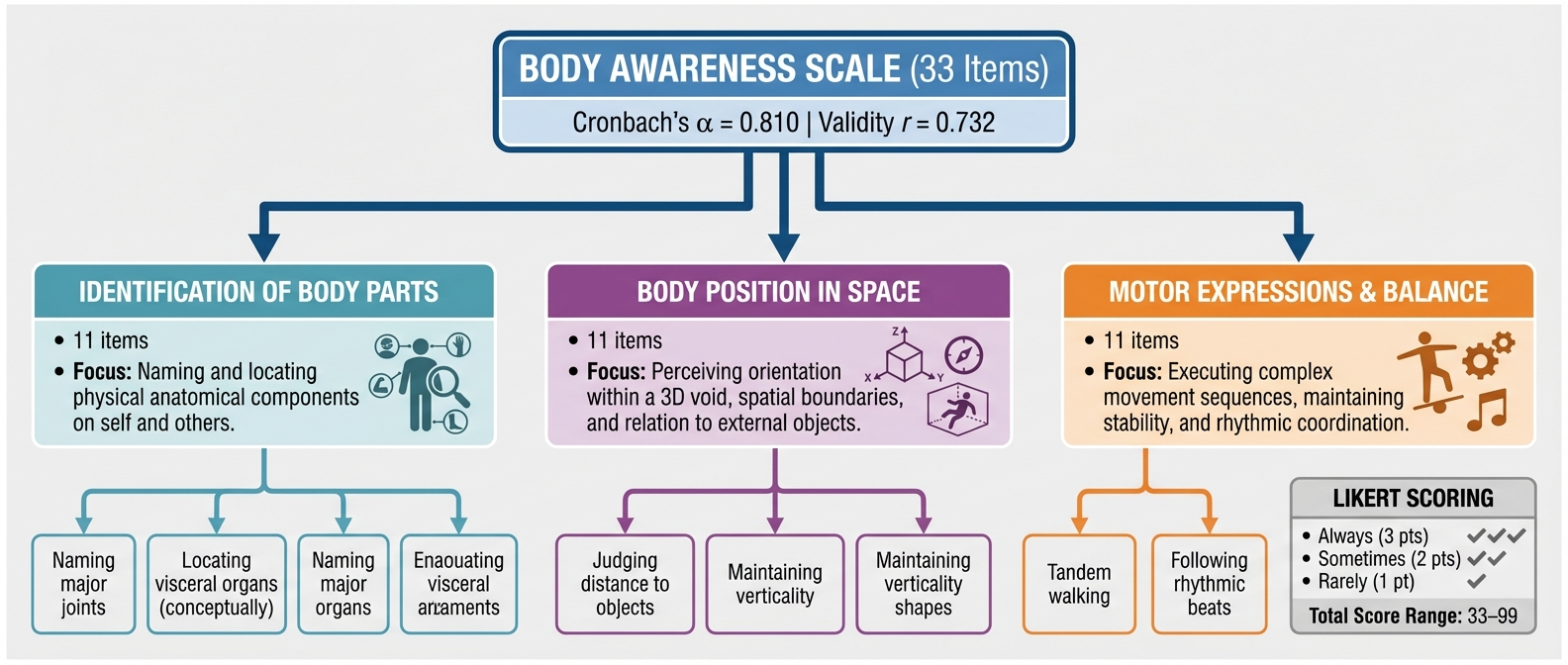}
  \caption{The assessment framework adopted in the proposed research}
  \label{fig:assess_framework}
\end{figure}

The assessment of the study’s variables relied on three primary standardized and researcher-prepared instruments. First, the Stanford-Binet Intelligence Scale (5th Edition), as adapted by Abu El-Niel \cite{abu_el_nil_2011_sb5}, was individualy administered to ensure that all participants possessed the cognitive capacity to engage with the VR stimuli. Second, the Conners Rating Scale for Teachers (CTRS-28/39) \cite{el_beheiry_2021_conners}, served as the diagnostic benchmark for ADHD symptoms, specifically measuring hyperactivity, impulsivity, and inattention through a 36-item, three-point Likert scale. Finally, the researcher developed a specialized Body Awareness Scale for children with ADHD, which underwent rigorous validation. This scale consists of 33 items categorized into three sub-dimensions: identification of body parts, perception of body position in space, and motor expressions. The scale demonstrated high internal consistency (Cronbach’s $\alpha = 0.810$) and significant concurrent validity ($r = 0.732$) when compared with established psychomotor assessments. The assessment framework that is adopted in this study is visually summarized in Fig.~\ref{fig:assess_framework}.

\subsection{The Virtual Reality Intervention Program}
The core of the methodology involved a structured VR training program designed to simulate a "quasi-realistic" interactive environment. The program was implemented over a 10-week period from January 1 to April 1, 2025, totaling 36 sessions. Each session was designed to provide immersive 3D stimuli that required the children to navigate virtual spaces, balance on simulated structures, and coordinate their movements in response to sensory feedback. By engaging the children in these interactive "embodied" tasks, the program aimed to refine the internal "Body Schema" and reduce the spatial neglect often associated with ADHD. Following the conclusion of the 36 sessions, post-tests were administered, followed by a one-month follow-up assessment to determine the durability of the psychomotor improvements.

\section{Results and Discussion}
\subsection{Baseline Group Homogeneity}
To establish the validity of the experimental intervention, the researcher first conducted a comparative analysis of the experimental and control groups at the baseline stage. As detailed in Table 1, the Mann-Whitney U test confirmed that there were no statistically significant differences ($p > 0.05$) between the two groups across any of the controlled variables, including chronological age, intelligence quotient (IQ), socioeconomic status, and initial ADHD symptom severity. Most importantly, the pre-test scores on the Body Awareness Scale showed no significant variance ($p = 0.670$), ensuring that any subsequent improvements could be attributed specifically to the virtual reality program.

\begin{table}[htbp]
    \centering
    \caption{Statistical Comparison of Experimental and Control Groups at Baseline ($N=10$)}
    \label{tab:baseline_comparison}
    \begin{tabular}{lcccccc}
        \toprule
        \textbf{Variable} & \textbf{Group} & \textbf{Mean} & \textbf{Std. Dev} & \textbf{Mean Rank} & \textbf{Z-Value} & \textbf{Sig. (p)} \\
        \midrule
        \textbf{Chronological Age} & Exp. / Cont. & 62.00 / 62.40 & 0.00 / 0.548 & 4.50 / 6.50 & -1.50 & 0.134 \\
        \addlinespace
        \textbf{IQ (Stanford-Binet)} & Exp. / Cont. & 103.8 / 105.4 & 2.59 / 2.89 & 4.80 / 6.20 & -0.736 & 0.426 \\
        \addlinespace
        \textbf{Socioeconomic Status} & Exp. / Cont. & 39.2 / 41.2 & 1.92 / 5.40 & 4.50 / 6.50 & -1.048 & 0.295 \\
        \addlinespace
        \textbf{ADHD (Conners Scale)} & Exp. / Cont. & 42.20 / 42.00 & 6.50 / 6.16 & 5.60 / 5.40 & 0.104 & 0.917 \\
        \addlinespace
        \textbf{Body Awareness (Pre)} & Exp. / Cont. & 35.80 / 35.00 & 1.92 / 2.12 & 5.90 / 5.10 & 0.426 & 0.670 \\
        \bottomrule
    \end{tabular}
\end{table}

\subsection{Analysis of Program Efficacy}
The primary research question sought to determine the effectiveness of the Virtual Reality (VR) training program in improving body awareness. Following the 36-session intervention, the Wilcoxon Signed-Ranks Test was utilized to compare the pre-test and post-test scores of the experimental group. As demonstrated in Table 2, the results indicated statistically significant differences in favor of the post-test measurement across all assessed sub-dimensions: the identification of body parts, perception of the body's position in space, and the execution of complex motor expressions. Notably, the total body awareness mean rank rose from a baseline of 5.90 to a significantly higher level, confirming the program's robust efficacy.

The researcher attributes these gains to the immersive, "quasi-realistic" nature of the VR environment. Unlike traditional 2D media, which can reduce motivation in children with ADHD, VR provides a multisensory feedback loop that reinforces the child's internal "Body Schema". By requiring the child to physically navigate a 3D space to achieve goals, the program directly addresses the "spatial neglect" and motor planning deficits typical of this population.

\begin{table}[htbp]
    \centering
    \caption{Wilcoxon Signed-Ranks Test Comparison of Pre-test and Post-test Scores for the Experimental Group ($n=5$)}
    \label{tab:efficacy_results}
    \begin{tabular}{lcccc}
        \toprule
        \textbf{Assessment Dimension} & \textbf{Pre-test Rank} & \textbf{Post-test Rank} & \textbf{Sig. (p)} & \textbf{Direction} \\
        \midrule
        Body Parts Identification & Low & High & $p < 0.05$ & Post-test \\
        Body Position in Space & Low & High & $p < 0.05$ & Post-test \\
        Motor Expressions & Low & High & $p < 0.05$ & Post-test \\
        \addlinespace
        \textbf{Total Body Awareness} & \textbf{5.90} & \textbf{Significant Increase} & \boldmath$p < 0.05$ & \textbf{Post-test} \\
        \bottomrule
    \end{tabular}
    \caption*{Note: All statistical differences were significant in favor of the post-test measurement.}
\end{table}

\subsection{Sustainability and Continuity of Effect}
A secondary objective was to examine the sustainability of the program’s effectiveness after the cessation of the intervention. A follow-up assessment conducted one month after the post-test revealed no statistically significant differences in the participants' psychomotor performance. This finding, quantified in Table 3, confirms the continuity of the program’s impact by showing that the "High" mean ranks achieved during the post-test remained stable ($p > 0.05$) during the follow-up period.

The stability of these scores indicates that the VR intervention facilitates long-term neuroplasticity rather than temporary behavioral change. By providing a safe and engaging environment for repetitive spatial challenges, the framework allows children to develop a robust sense of self-awareness and physical control that persists beyond the immediate therapeutic setting. This longitudinal retention suggests that the 36-session immersive training fostered a more permanent "Body Schema" in the participants.

\begin{table}[htbp]
    \centering
    \caption{Sustainability of Program Effect: Comparison of Post-test and Follow-up Scores ($n=5$)}
    \label{tab:stability_results}
    \begin{tabular}{lcccc}
        \toprule
        \textbf{Measurement Phase} & \textbf{Mean Rank} & \textbf{Z-Value} & \textbf{Sig. (p)} & \textbf{Result} \\
        \midrule
        Post-test & High & -- & -- & -- \\
        Follow-up (One Month) & High & -- & $p > 0.05$ & Stable \\
        \bottomrule
    \end{tabular}
    \caption*{Note: No statistically significant differences were found between the post-test and follow-up measurements, indicating the continuity of the program's effect.}
\end{table}

\subsection{Limitations}
The findings suggest that the integration of "Embodied Technology" represents a significant advancement over traditional sedentary interventions. The high levels of motivation observed during the program acted as a catalyst for cognitive-motor synchronization. However, as noted in the research, certain limitations exist: the small cohort size ($N=10$) and the specific focus on early childhood (ages 4–7) necessitate caution regarding the broad generalizability of these results to older populations or different cultural settings. Future research should investigate the scalability of this framework using more accessible VR platforms to reach a wider demographic of neurodivergent children.

\section{Conclusion}
In this research, we proposed and evaluated an immersive training framework based on Virtual Reality (VR) technology specifically designed to address body awareness deficits in children with Attention Deficit Hyperactivity Disorder (ADHD). This work fills a significant gap in the current therapeutic landscape, where traditional methods often fail to engage children with ADHD or provide the precise, simulated spatial environments necessary to correct psychomotor imbalances. By shifting from passive screen-based observation to active, "quasi-realistic" physical participation, this study provides a novel evidence-based approach to early childhood intervention in the Arabic educational context.   

The proposed framework was structured as a multi-dimensional intervention focusing on the identification of body parts, the perception of the body within a spatial void, and the execution of complex motor expressions. By utilizing interactive 3D environments, the system provided immediate sensory feedback that allowed participants to map their physical boundaries more accurately. This approach moved beyond simple motor drills, integrating cognitive-motor synchronization to help children manage their movements with greater intent and less impulsivity.   

The results of the three-month intervention provided deep insights into the intersection of technology and neurodevelopmental therapy. While the data showed a marked increase in body awareness scores for the experimental group, the deeper insight lies in the stability of these gains during follow-up assessments. This suggests that the VR-based framework facilitates lasting psychomotor adaptation rather than mere temporary behavioral compliance. Furthermore, the high level of engagement observed during the 36 sessions indicates that the gamified nature of VR acts as a powerful motivator, effectively bypassing the attention barriers typically associated with ADHD.   

This framework holds immense potential for scalability, offering a safe and controlled environment for children to practice high-stakes spatial navigation tasks without the risk of physical injury. However, the approach is not without its challenges. The reliance on specialized hardware and the small, age-specific cohort of children aged 4 to 7 years present limitations in terms of immediate generalizability and accessibility. Moving forward, future research should focus on longitudinal studies with larger, more diverse samples to explore how these VR-enhanced psychomotor skills translate into long-term academic and social success. We also suggest investigating the integration of more accessible VR platforms to ensure these benefits can reach a wider range of clinical and educational settings.

\section*{Acknowledgments}
During the preparation of this work, the authors used ChatGPT of OpenAI and Gemini of Google for language refinement and paraphrasing. We also adopted PaperBanana\footnote{\url{https://paper-banana.org/}} to enhance the design of Fig.~\ref{fig:exp_timeline} and Fig.~\ref{fig:assess_framework}. All intellectual contributions, critical analysis, and final edits were conducted by the authors. After using the aforementioned tools/services, the authors reviewed and edited the content as needed and take full responsibility for the content of the published article.

\bibliographystyle{unsrt}  
\bibliography{references}

\end{document}